\documentclass{llncs}

\usepackage{latexsym,amssymb}
\usepackage{amsmath}
\usepackage{stmaryrd}

\newbox\tempa
\newbox\tempb
\newdimen\tempc
\def\mud#1{\hfil $\displaystyle{\mathstrut #1}$\hfil}
\def\rig#1{\hfil $\displaystyle{#1}$}
\def\irulehelp#1#2#3{\setbox\tempa=\hbox{$\displaystyle{\mathstrut #2}$}%
                        \setbox\tempb=\vbox{\halign{##\cr
        \mud{#1}\cr
        \noalign{\vskip\the\lineskip}%
        \noalign{\hrule height 0pt}%
        \rig{\vbox to 0pt{\vss\hbox to 0pt{${\; #3}$\hss}\vss}}\cr
        \noalign{\hrule}%
        \noalign{\vskip\the\lineskip}%

        \mud{\copy\tempa}\cr}}%
                      \tempc=\wd\tempb
                      \advance\tempc by \wd\tempa
                      \divide\tempc by 2 }
\def\irule#1#2#3{{\irulehelp{#1}{#2}{#3}%
                     \hbox to \wd\tempa{\hss \box\tempb \hss}}}

\newcommand{\lra}{\longrightarrow}
\newcommand{\ra}{\rightarrow}

\newcommand{\ex}{\exists}
\newcommand{\fa}{\forall}
\newcommand{\dotfa}{\dot{\fa}}
\newcommand{\dotvee}{\dot{\vee}}
\newcommand{\dotneg}{\dot{\neg}}
\def\nulll{\mbox{\it Null\/}}
\def\pred{\mbox{\it Pred\/}}
\def\succ{\mbox{\it Succ\/}}
 
\begin{document}     
\title{Simple Type Theory as a Clausal Theory}
\author{Gilles Dowek}
\date{}
\institute{\'Ecole polytechnique and INRIA\\
LIX, \'Ecole polytechnique,
91128 Palaiseau Cedex, France. \\
{\tt Gilles.Dowek@polytechnique.edu, http://www.lix.polytechnique.fr/\~{}dowek}}
\maketitle

\thispagestyle{empty}

\section{Introduction}

Deduction modulo is an extension of first-order predicate logic where
axioms are replaced by rewrite rules. For instance, the axiom
$P \Leftrightarrow (Q \Rightarrow R)$ is replaced by the rule $P \lra (Q
\Rightarrow R)$. These rules define an equivalence relation and, in a
proof, a proposition can be replaced by an equivalent one at any time.
A presentation of Simple Type Theory in Deduction modulo, called
$HOL$, has been given in \cite{DHKHOL}. 

{\em Polarized deduction modulo} \cite{stacs} is an extension of Deduction
modulo where possibly different rewrite rules apply to the
negative and positive occurrences of atomic propositions. 
Like any theory expressed in Deduction modulo, Simple Type Theory 
can be expressed in Polarized deduction modulo. Each rule just needs
to be duplicated in a negative and a positive instance. 

A rewrite system in Polarized deduction modulo is said to be clausal
when negative rules rewrite atomic propositions to clausal
propositions and positive rules rewrite atomic propositions to
the negation of clausal propositions. This way, clauses rewrite to
clauses, which is a useful property in automated theorem proving
\cite{polar}. 

We give in this note a presentation of Simple Type Theory as a clausal
rewrite system in Polarized deduction modulo, called $HOL^{\pm}$. 

This system is does not have the cut elimination property in general
but cut elimination holds for sequents well-formed in the language of
$HOL$ and, for such sequents, provability in $HOL^{\pm}$ and in $HOL$
are equivalent.

\section{Polarized deduction modulo}

\begin{definition}[Polarized rewrite system]
A {\em polarized rewrite system} is a triple ${\cal R} = \langle {\cal
E}, {\cal R}_-, {\cal R}_+ \rangle$ where ${\cal E}$ is a set of
equations between terms, ${\cal R}_-$ and ${\cal R}_+$ are sets of
rewrite rules whose left hand sides are atomic propositions and right
hand sides are arbitrary propositions.  The rules of ${\cal R}_-$ are
called {\em negative} and those of ${\cal R}_+$ are called {\em
positive}.
\end{definition}

\begin{definition}[Polarized rewriting]
Let 
${\cal R} = \langle {\cal E}, {\cal R}_-, {\cal R}_+ \rangle$ 
be a polarized rewrite system. 
We define the equivalence relation $=_{\cal E}$ as the congruence on terms 
generated by the equations of ${\cal E}$.
We then define the one step negative and positive rewriting
relations $\lra_-$ and $\lra_+$ as follows.
\begin{itemize}
\item
If $t_i =_{\cal E} t'$ then 
$P(t_1, \ldots, t_i, \ldots, t_n) \lra_-
P(t_1, \ldots, t', \ldots, t_n)$\\
and $P(t_1, \ldots, t_i, \ldots, t_n) \lra_+
P(t_1, \ldots, t', \ldots, t_n)$. 

\item
If $P \lra A$ is a rule of ${\cal R}_-$ and $\sigma$ is a substitution
then $\sigma P \lra_- \sigma A$.

If $P \lra A$ is a rule of ${\cal R}_+$ and $\sigma$ is a substitution
then $\sigma P \lra_+ \sigma A$.

\item 
If $A \lra_+ A'$ then $\neg A \lra_- \neg A'$. 
If $A \lra_- A'$ then $\neg A \lra_+ \neg A'$.

\item
If ($A \lra_- A'$ and $B = B'$)
or 
($A = A'$ and $B \lra_- B'$),
then\\
$A \wedge B \lra_- A' \wedge B'$ and $A \vee B \lra_- A' \vee B'$. 

If ($A \lra_+ A'$ and $B = B'$)
or 
($A = A'$ and $B \lra_+ B'$),
then\\
$A \wedge B \lra_+ A' \wedge B'$ and $A \vee B \lra_+ A' \vee B'$.

\item 
If ($A \lra_+ A'$ and $B = B'$)
or 
($A = A'$ and $B \lra_- B'$),
then\\
 $A \Rightarrow B \lra_- A' \Rightarrow B'$.

If ($A \lra_- A'$ and $B = B'$)
or 
($A = A'$ and $B \lra_+ B'$),
then\\
 $A \Rightarrow B \lra_+ A' \Rightarrow B'$.

\item 
If $A \lra_- A'$ then $\fa x~A \lra_- \fa x~A'$ 
and $\ex x~A \lra_- \ex x~A'$. 

If $A \lra_+ A'$ then $\fa x~A \lra_+ \fa x~A'$ 
and $\ex x~A \lra_+ \ex x~A'$. 
\end{itemize}

We define the sequent one step term rewriting relation $\lra$ as
follows.
\begin{itemize}
\item If $A \lra_- A'$ then $(\Gamma, A \vdash \Delta) \lra
(\Gamma, A' \vdash \Delta)$.
\item If $A \lra_+ A'$ then $(\Gamma \vdash A, \Delta) \lra
(\Gamma \vdash A', \Delta)$.
\end{itemize}
\end{definition}

\begin{figure}
\noindent\framebox{\parbox{\textwidth
}{
$$
\hspace*{-4cm}
\begin{array}{c}
\irule{}
      {A \vdash B}
      {\mbox{axiom if $A \lra_-^* P, B \lra_+^* P$ and $P$ atomic}}
\vspace{1.5mm}\\
\irule{\Gamma, B \vdash \Delta ~~~ \Gamma \vdash C, \Delta}
      {\Gamma \vdash \Delta}
      {\mbox{cut if $A \lra_-^* B, A \lra_+^* C$}}
\vspace{1.5mm}\\
\irule{\Gamma, B, C \vdash \Delta}
      {\Gamma, A \vdash \Delta}
      {\mbox{contr-left if $A \lra_-^* B, A \lra_-^* C$}}
\vspace{1.5mm}\\
\irule{\Gamma \vdash B,C,\Delta}
      {\Gamma \vdash A,\Delta}
      {\mbox{contr-right if $A \lra_+^* B, A \lra_+^* C$}}
\vspace{1.5mm}\\
\irule{\Gamma \vdash \Delta}
      {\Gamma, A \vdash \Delta}
      {\mbox{weak-left}}
\vspace{1.5mm}\\
\irule{\Gamma \vdash\Delta}
      {\Gamma \vdash A,\Delta}
      {\mbox{weak-right}}
\vspace{1.5mm}\\
\irule{}
      {\Gamma \vdash A, \Delta}
      {\mbox{$\top$-right if $A \lra_+^* \top$}}
\vspace{1.5mm}\\
\irule{}
      {\Gamma, A \vdash \Delta}
      {\mbox{$\bot$-left if $A \lra_-^* \bot$}}
\vspace{1.5mm}\\
\irule{\Gamma \vdash B, \Delta}
      {\Gamma, A \vdash  \Delta}
      {\mbox{$\neg$-left if $A \lra_-^* \neg B$}}
\vspace{1.5mm}\\
\irule{\Gamma, B \vdash \Delta}
      {\Gamma \vdash  A, \Delta}
      {\mbox{$\neg$-right if $A \lra_+^* \neg B$}}
\vspace{1.5mm}\\
\irule{\Gamma, B, C \vdash \Delta}
      {\Gamma, A \vdash  \Delta}
      {\mbox{$\wedge$-left if $A \lra_-^* (B \wedge C)$}}
\vspace{1.5mm}\\
\irule{\Gamma \vdash B, \Delta~~~\Gamma \vdash C, \Delta}
      {\Gamma \vdash A, \Delta}
      {\mbox{$\wedge$-right if $A \lra_+^* (B \wedge C)$}}
\vspace{1.5mm}\\
\irule{\Gamma, B \vdash \Delta ~~~ \Gamma, C \vdash \Delta}
      {\Gamma, A \vdash  \Delta}
      {\mbox{$\vee$-left if $A \lra_-^* (B \vee C)$}}
\vspace{1.5mm}\\
\irule{\Gamma \vdash B, C, \Delta}
      {\Gamma \vdash A, \Delta}
      {\mbox{$\vee$-right if $A \lra_+^* (B \vee C)$}}
\vspace{1.5mm}\\
\irule{\Gamma \vdash B, \Delta ~~~ \Gamma, C \vdash \Delta}
      {\Gamma, A \vdash  \Delta}
      {\mbox{$\Rightarrow$-left if $A \lra_-^* (B \Rightarrow C)$}}
\vspace{1.5mm}\\
\irule{\Gamma, B \vdash C, \Delta}
      {\Gamma \vdash  A, \Delta}
      {\mbox{$\Rightarrow$-right if $A \lra_+^* (B \Rightarrow C)$}}
\vspace{1.5mm}\\
\irule{\Gamma, C \vdash \Delta}
      {\Gamma, A \vdash  \Delta}
      {\mbox{$\langle x, B, t \rangle$ $\fa$-left if $A \lra_-^* \fa x~B$, $(t/x)B \lra_-^* C$}}
\vspace{1.5mm}\\
\irule{\Gamma \vdash B, \Delta}
      {\Gamma \vdash  A, \Delta}
      {\mbox{$\langle x, B \rangle$ $\fa$-right 
   if $A \lra_+^* \fa x~B$, $x \not\in FV(\Gamma \Delta)$}}
\vspace{1.5mm}\\
\irule{\Gamma, B \vdash \Delta}
      {\Gamma, A \vdash  \Delta}
      {\mbox{$\langle x, B \rangle$ $\ex$-left
   if $A \lra_-^* \ex x~B$, $x \not\in FV(\Gamma \Delta)$}}
\vspace{1.5mm}\\
\irule{\Gamma \vdash C, \Delta}
      {\Gamma \vdash A, \Delta}
      {\mbox{$\langle x, B, t \rangle$ $\ex$-right if $A \lra_+^* \ex x~B$, $(t/x)B \lra_+^* C$}}
\end{array}
$$
\caption{Polarized sequent calculus modulo\label{sequent}}
}}
\end{figure}

As usual, if $R$ is any binary relation, we write $R^*$ for 
its reflexive-transitive closure.
The rules of {\em Polarized sequent calculus modulo} are those of
Fig.~\ref{sequent}. 
Proof checking is decidable when the relations 
$\lra_-^*$ and $\lra_+^*$ are. 
The usual, non polarized, 
Deduction modulo can be recovered by taking 
${\cal R}_- = {\cal R}_+$ and 
predicate logic by taking 
${\cal E} = {\cal R}_- = {\cal R}_+ = \varnothing$. 

A {\em theory} is a pair $\langle {\cal R}, {\cal T}\rangle$ formed
with a polarized rewrite system ${\cal R}$ and a set of axioms ${\cal T}$.
We say that the sequent $\Gamma \vdash \Delta$ is
{\em provable in the theory $\langle {\cal R}, {\cal T} \rangle$}, or that
it is {\em provable in ${\cal T}$ modulo ${\cal R}$}, 
if there exists
a finite subset ${\cal T}'$ of ${\cal T}$ such that the sequent 
$\Gamma, {\cal T}' \vdash \Delta$ is provable in Polarized sequent 
calculus modulo ${\cal R}$.
When ${\cal T}$ is empty, we simply say that the sequent $\Gamma  \vdash
\Delta$ is {\em provable modulo ${\cal R}$}.
When ${\cal R}$ is empty, we say that the sequent $\Gamma  \vdash
\Delta$ is {\em provable in ${\cal T}$ in predicate logic}.

As discussed in \cite{stacs}, rewriting, in general, has two
properties. First, it is oriented and, for instance, the proposition
$x \in {\cal P}(y)$ rewrites to $\fa z~(z \in x \Rightarrow z \in y)$, 
but $\fa z~(z \in x \Rightarrow z \in y)$ does not rewrite to 
$x \in {\cal P}(y)$.  Then, rewriting
preserves provability. For instance, the proposition 
$x \in {\cal P}(y)$ rewrites to $\fa z~(z \in x \Rightarrow z \in y)$ 
that is provably equivalent.  Thus, we can always transform the proposition 
$x \in {\cal P}(y)$ into $\fa z~(z \in x \Rightarrow z \in y)$ and we never 
need to backtrack on this operation.  When rewriting is polarized, the 
first property is
kept, but not the second. For instance, if we have the negative rule
$P \lra Q$, the sequent $P \vdash P$ can be proved with the axiom
rule, but its normal form $Q \vdash P$ cannot.

\begin{definition}[Literal, Clausal proposition]
A proposition is 
a {\em literal} if it is either atomic of the negation
of an atomic proposition.
A proposition is {\em clausal} if it is $\bot$ or of the form 
$\fa x_1 \ldots \fa x_p~(L_1 \vee \ldots \vee L_n)$ where 
$L_1, \ldots, L_n$ are literals and $x_1, \ldots, x_p$ variables. 
\end{definition}

\begin{definition}[Clausal rewrite system]
A rewrite system is {\em clausal} if negative rules
rewrite atomic propositions to a clausal propositions and 
positive rules atomic propositions to negations of clausal
propositions.
\end{definition}

\section{Equivalence} 

We want to show that rewrite rules build-in axioms, i.e.  that for
each rewrite system ${\cal R}$, there is a set of axioms ${\cal
U}_{\cal R}$ such that a sequent is provable modulo
${\cal R}$ if and only if it is provable in ${\cal U}_{\cal R}$ in
predicate logic. The set of axioms we wish to consider contains for
each equational axiom $t = u$ of ${\cal E}$, the universal closure of the
proposition $t = u$, for each rule $P \lra A$ of ${\cal R}_-$, 
the universal closure of the proposition $P \Rightarrow A$, 
and for each rule $P \lra A$ of ${\cal R}_+$, the 
universal closure of the
proposition $A \Rightarrow P$. A problem is that the language we
start with need not contain an equality predicate. Thus, we must first
add such a predicate and the axioms of equality and prove that this
extension is conservative.

\begin{definition}[Compatibility]
Let ${\cal R}$ be a polarized rewrite system and ${\cal T}$ and 
${\cal U}$ be two sets of axioms.
The theory $\langle {\cal R}, {\cal T} \rangle$ is {\em compatible} with 
${\cal U}$ if
\begin{itemize}
\item if $A \lra_-^* B$ in ${\cal R}$, then 
$\vdash A \Rightarrow B$ is provable in ${\cal U}$ in predicate logic,
\item if $A \lra_+^* B$ in ${\cal R}$, then 
$\vdash B \Rightarrow A$ is provable in ${\cal U}$ in predicate logic,
\item if $A \in {\cal T}$, then $\vdash A$ is provable in ${\cal U}$
in predicate logic, 
\item if $A \in {\cal U}$, then $\vdash A$ is provable in ${\cal T}$ modulo 
${\cal R}$.
\end{itemize}
\end{definition}

\begin{proposition}[Equivalence]
\label{equivalence}
Let ${\cal R}$ be a polarized rewrite system and ${\cal T}$ and ${\cal
U}$ be two sets of axioms such that the theory $\langle {\cal R},
{\cal T} \rangle$ is compatible with ${\cal U}$, then a sequent is
provable in ${\cal T}$ modulo ${\cal R}$, if and only if it 
is provable in ${\cal U}$ in predicate logic.
\end{proposition}

\proof{If the sequent $\Gamma \vdash \Delta$ is provable in 
${\cal U}$ in predicate logic,
there exists a finite subset ${\cal U}'$ of ${\cal U}$ such that 
$\Gamma, {\cal U}' \vdash \Delta$ is provable in predicate logic and hence 
modulo ${\cal R}$. 
Each $U_i$ in ${\cal U}'$ is provable in ${\cal
T}$ modulo ${\cal R}$, thus, for each $U_i$, there exists a finite
subset ${\cal T}'_i$ of ${\cal T}$ such that ${\cal T}'_i \vdash U_i$
is provable modulo ${\cal R}$. Let ${\cal T}'$ be the union of all the 
${\cal T}_i$'s. Using the cut rule, we build a proof of $\Gamma, {\cal T}'
\vdash \Delta$ modulo ${\cal R}$. Thus, the sequent $\Gamma \vdash
\Delta$ is provable in ${\cal T}$ modulo ${\cal R}$. 

The converse is a simple induction over proof structure.}

\begin{definition}
Let ${\cal R}$ be a polarized rewrite
system. Let ${\cal A}_{\cal R}$ be 
the set of axioms containing
\begin{itemize}
\item for each pair of propositions $A$ and $B$ such that 
$A \lra_-^* B$, the universal closure of $A \Rightarrow B$, 
\item for each pair of propositions $A$ and $B$ such that 
$A \lra_+^* B$, the universal closure
of $B \Rightarrow A$.
\end{itemize}
\end{definition}

\begin{proposition}\label{AR}
Let ${\cal R}$ be a polarized rewrite system and ${\cal T}$ be a set
of axioms.
Then, the theory $\langle {\cal R}, {\cal T} \rangle$ and the set of axioms
${\cal A}_{\cal R} \cup {\cal T}$ are compatible.
\end{proposition}

\proof{If $A \lra_-^* B$ in ${\cal R}$, then the universal closure of 
$A \Rightarrow B$ is an element of ${\cal A}_{\cal R}$. Thus,
the sequent $\vdash A \Rightarrow B$ is provable in ${\cal A}_{\cal R} \cup {\cal
    T}$
in predicate logic.
If $A \lra_+^* B$ in ${\cal R}$, then the universal closure of 
$B \Rightarrow A$ is an element of ${\cal A}_{\cal R}$. Thus, the sequent
$\vdash B \Rightarrow A$ is provable in ${\cal A}_{\cal R} \cup {\cal
T}$ in predicate logic.
If $A \in {\cal T}$, then $A \in {\cal A}_{\cal R} \cup {\cal T}$ 
and thus the sequent 
$\vdash A$ is provable in 
${\cal A}_{\cal R} \cup {\cal T}$ in predicate logic.

Conversely, if $A \in {\cal T}$, then the sequent $\vdash A$ is
provable in ${\cal T}$ modulo ${\cal R}$ and 
if $A \in {\cal A}_{\cal R}$, then the sequent $\vdash A$ is
provable in ${\cal T}$ modulo ${\cal R}$.}

\begin{definition}[Model]
Let ${\cal R}$ be a polarized rewrite system and ${\cal T}$ be a
set of axioms, a {\em model} of the theory 
$\langle {\cal R}, {\cal T} \rangle$ is 
a model of the set of axioms ${\cal A}_{\cal R} \cup {\cal T}$. 
\end{definition}

\begin{proposition}[Soundness and completeness]
A sequent $\Gamma \vdash \Delta$ is provable in ${\cal T}$ modulo 
${\cal R}$ if and only if valid in all models of $\langle {\cal R},
{\cal T} \rangle$. 
\end{proposition}

\proof{By Propositions \ref{equivalence} and \ref{AR}, the sequent 
$\Gamma \vdash \Delta$ is provable in ${\cal T}$ modulo ${\cal R}$ 
if and only if it is provable in ${\cal A}_{\cal R} \cup {\cal T}$. 
By the soundness and completeness theorem of predicate logic it is 
provable in ${\cal A}_{\cal R} \cup {\cal T}$ if and only if it is
valid in all models of  ${\cal A}_{\cal R} \cup {\cal T}$, i.e. in all
models of $\langle {\cal R}, {\cal T} \rangle$.}

\begin{definition}[Equality model]
Let ${\cal R}$ be a polarized rewrite system and ${\cal T}$ be a
set of axioms. An {\em equality model} of $\langle {\cal R}, {\cal T}
\rangle$ is a model where if $t =_{\cal E} u$ then for all $\phi$, 
$\llbracket t \rrbracket_{\phi} = \llbracket u \rrbracket_{\phi}$. 
\end{definition}

\begin{proposition}[Soundness and completeness for equality models]
A sequent $\Gamma \vdash \Delta$ is provable in ${\cal T}$ modulo 
${\cal R}$ if and only if valid in all equality models of $\langle {\cal R},
{\cal T} \rangle$. 
\end{proposition}

\proof{All we need to prove is that for each model ${\cal M}$ of 
$\langle {\cal R}, {\cal T} \rangle$ we can build an equality model of
$\langle {\cal R}, {\cal T} \rangle$. 
Let ${\cal M}$ be a model of $\langle {\cal R}, {\cal T} \rangle$.
We write ${\cal M}_T$ for the domain of 
${\cal M}$ of sort $T$, $\hat{f}$ for the interpretation
of the function symbol $f$ and $\hat{P}$ for the interpretation
of the predicate symbol $P$.
For each sort $T$, we define the 
relation $\sim_T$ on the elements of ${\cal M}_T$, by
$a \sim_T b$ if and only if there
exists two terms $t$ and $u$ of sort $T$ and a valuation $\phi$ such that 
$t =_{\cal E} u$, 
$a = \llbracket t \rrbracket_{\phi}$ and 
$b = \llbracket u \rrbracket_{\phi}$. This relation is obviously an
equivalence relation and it is compatible with the interpretation of
all the function symbols. To prove that it is compatible with the denotation
of the predicate
symbols, we remark that if $t =_{\cal E} u$ then 
$P(t_1, ..., t, ..., t_n) \lra_- P(t_1, ..., u, ..., t_n)$ and 
$P(t_1, ..., t, ..., t_n) \lra_+ P(t_1, ..., u, ..., t_n)$, thus
the proposition 
$P(t_1, ..., t, ..., t_n) \Leftrightarrow P(t_1, ..., u, ..., t_n)$ 
is provable modulo ${\cal R}$ and thus valid in ${\cal M}$.
We finally define a model ${\cal M}'$ by taking 
${\cal M}'_T = {\cal M}_T / \equiv_T$
and by interpreting the function symbol $f$ by the 
function $\hat{f}/\equiv$ and the predicate symbol $f$ by the function
$\hat{P}/\equiv$. The propositions valid in the models ${\cal M}$ and 
${\cal M}'$ are the same.}

\begin{definition}
Let ${\cal L}$ be a language containing an equality predicate in each
sort. Let 
${\cal R}$ be a polarized rewrite system in ${\cal L}$. 
Let ${\cal U}_{\cal R}$ be 
the set of axioms containing
\begin{itemize}
\item the axioms of equality for ${\cal L}$, 
\item for each equational axiom $t = u$ of ${\cal E}$, the universal closure
of the proposition $t = u$, 
\item for each rule $P \lra A$ of ${\cal R}_-$, the universal closure
of the proposition $P \Rightarrow A$, 
\item for each rule $P \lra A$ of ${\cal R}_+$, the universal closure
of the proposition $A \Rightarrow P$.
\end{itemize}
\end{definition}

\begin{proposition}\label{UR}
Let ${\cal L}$ be a language containing an equality predicate in each
sort. Let ${\cal E}q$ be the axioms of equality for ${\cal L}$. 
Let ${\cal R}$ be a polarized rewrite system in ${\cal L}$. 
Then, the theory $\langle {\cal R}, {\cal E}q \rangle$ and the set of axioms
${\cal U}_{\cal R}$ are compatible.
\end{proposition}

\proof{It is routine to check that if $A \lra_-^* B$ in ${\cal R}$, then 
the sequent $\vdash A \Rightarrow B$ is provable in 
${\cal U}_{\cal R}$ in predicate logic,
and if $A \lra_+^* B$ in ${\cal R}$, then 
the sequent $\vdash B \Rightarrow A$ is provable in 
${\cal U}_{\cal R}$
in predicate logic.
If $A$ is an axiom of ${\cal E}q$, then it is an axiom of ${\cal U}_{\cal R}$,
hence the sequent $\vdash A$ is provable in ${\cal U}_{\cal R}$ in 
predicate logic.

Conversely, we check, considering each of the four cases, that if $A
\in {\cal U}_{\cal R}$, then the sequent $\vdash A$ is provable in
${\cal E}q$ modulo ${\cal R}$.}

\begin{proposition}\label{conservative}
Let ${\cal R}$ be a polarized rewrite system in a language ${\cal
L}$. Let ${\cal L}'$ be the language obtained by adding an equality 
symbol in each sort of ${\cal L}$. Let ${\cal E}q$ be the axioms of 
equality for ${\cal L}'$. Then, the theory $\langle {\cal R}, {\cal
  E}q \rangle$ is a conservative
extension of ${\cal R}$, i.e. a sequent $\Gamma \vdash \Delta$ of
${\cal L}$ is provable modulo ${\cal R}$ if and only if it is provable
in ${\cal E}q$ modulo ${\cal R}$.
\end{proposition}

\proof{An equality model of ${\cal R}$ extends to an equality model of
$\langle {\cal R}, {\cal E}q \rangle$ by interpreting equality by 
equality.}

\medskip

Remark that this proof would not go through if we did not consider equality
models. Indeed if $t =_{\cal E} u$, then $t = t \lra_- t = u$ 
and if $t = u$ were not valid in the model, it would not be a model of 
the proposition $t = t \Rightarrow t = u$.

\begin{proposition}\label{aaa}
Let ${\cal L}$ be a language and ${\cal R}$ be a polarized rewrite
system in ${\cal L}$. Let ${\cal L'}$ be the language obtained by
adding an equality symbol in each sort of ${\cal L}$. 
Then, a sequent $\Gamma \vdash \Delta$ of ${\cal L}$ is provable modulo
${\cal R}$ if and only if it is provable in ${\cal U}_{\cal R}$. 
\end{proposition}

\proof{Let ${\cal E}q$ be the axioms of equality for ${\cal L}'$. 
By Proposition \ref{conservative}, 
the sequent $\Gamma \vdash \Delta$ is provable modulo ${\cal
R}$ if and only if it is provable in ${\cal E}q$ modulo ${\cal R}$ and by
Propositions \ref{equivalence} and \ref{UR} it is provable 
in ${\cal E}q$ modulo ${\cal R}$ if and only if it is provable in 
${\cal U}_{\cal R}$.}

\section{Simple Type Theory as a clausal rewrite system}

A presentation of Simple Type Theory in non polarized deduction 
modulo has been given in \cite{DHKHOL}. 
To define it in polarized deduction modulo 
we just duplicate each rule. We also consider an extension of 
the system presented in \cite{DHKHOL} with rules expressing 
the existence of a non surjective injection $\succ$ of type 
$\iota \ra \iota$, that allow to prove the ``axiom'' of infinity.

\begin{definition}[The theory $HOL$]

The sorts are {\em simple types}, inductively defined by
\begin{itemize}
\item $\iota$ and $o$ are sorts,
\item if $T$ and $U$ are sorts then $T \ra U$ is a sort.
\end{itemize}
As usual, we write $T_1 \ra  \dots \ra T_n \ra U$ for 
$T_1 \ra  ( \dots \ra (T_n \ra U)\dots)$. 
The language contains
\begin{itemize}
\item for each pair of sorts $T, U$, a constant
$K_{T,U}$ of sort $T \ra U \ra T$,

\item for each triple of sorts $T, U, V$, a constant
$S_{T,U,V}$ of sort $(T \ra U \ra V) \ra (T \ra U) \ra T \ra V$,

\item a constant $\dot\vee$ or sort $o \ra o \ra o$,

\item a constant $\dot\neg$ or sort $o \ra o$, 

\item for each sort $T$, a constant $\dotfa_T$ of 
sort $(T \ra o) \ra o$, 

\item a constant $0$ of sort $\iota$, two constants $\succ$ and $\pred$ 
of sort $\iota \ra \iota$, and a constant $\nulll$ of sort $\iota \ra o$, 

\item for each pair of sorts $T, U$, a function symbol
$\alpha_{T,U}$ of rank $\langle T \ra U, T, U \rangle$, 

\item a predicate symbol $\varepsilon$ of rank $\langle o \rangle$.
\end{itemize}
As usual, we write $(t~u)$ for $\alpha_{T,U}(t,u)$ and
$(t~u_1~\ldots~u_n)$ for $(\ldots(t~u_1) \ldots u_n)$.
The rewrite rules are 
$$(K_{T,U}~x~y) =_{\cal E} x$$
$$(S_{T,U,V}~x~y~z) =_{\cal E} (x~z~(y~z))$$
$$(\pred~(\succ~x)) =_{\cal E} x$$
$$\begin{array}{rcl@{~}@{~}@{~}rcl}
\varepsilon(x~\dotvee~y) &\lra_-& (\varepsilon(x) \vee \varepsilon(y))&
\varepsilon(x~\dotvee~y) &\lra_+& (\varepsilon(x) \vee \varepsilon(y))\\
\varepsilon(\dotneg~x) &\lra_-& \neg \varepsilon(x)&
\varepsilon(\dotneg~x) &\lra_+& \neg \varepsilon(x)\\
\varepsilon(\dotfa_T~x) &\lra_-& \fa y~\varepsilon(x~y)&
\varepsilon(\dotfa_T~x) &\lra_+& \fa y~\varepsilon(x~y)\\
\varepsilon(\nulll~(S~x)) &\lra_-& \bot
& \varepsilon(\nulll~(S~x)) &\lra_+& \bot\\
\varepsilon(\nulll~0) &\lra_-& \top
&\varepsilon(\nulll~0) &\lra_+& \top\\
\end{array}$$
\end{definition}

The theory $HOL$ is not clausal. We now define a clausal 
theory $HOL^{\pm}$ and prove it is equivalent to $HOL$. 

\begin{definition}[The theory $HOL^{\pm}$]
The sorts are the same as those of $HOL$.
The symbols are the same as those of $HOL$ and, for each sort $T$, 
a function symbol $H_T$ of sort $(T \ra o) \ra T$. 
The rewrite rules are
$$(K_{T,U}~x~y) =_{\cal E} x$$
$$(S_{T,U,V}~x~y~z) =_{\cal E} (x~z~(y~z))$$
$$(\pred~(\succ~x)) =_{\cal E} x$$
$$\begin{array}{rcl@{~}@{~}@{~}rcl}
\varepsilon(x~\dotvee~y) &\lra_-& (\varepsilon(x) \vee \varepsilon(y))&
\varepsilon(x~\dotvee~y) &\lra_+& \neg \neg \varepsilon(x)\\
&&& \varepsilon(x~\dotvee~y) &\lra_+& \neg \neg \varepsilon(y)\\
\varepsilon(\dotneg~x) &\lra_-& \neg \varepsilon(x) &
\varepsilon(\dotneg~x) &\lra_+& \neg \varepsilon(x)\\
\varepsilon(\dotfa_T~x) &\lra_-& \fa y~\varepsilon(x~y) &
\varepsilon(\dotfa_T~x) &\lra_+& \neg \neg \varepsilon(x~H_T(x))\\
\varepsilon(\nulll~(S~x)) &\lra_-& \bot\\
&&&
\varepsilon(\nulll~0) &\lra_+& \neg \bot\\
\end{array}$$
\end{definition}

\begin{proposition}\label{sensdirect}
If a sequent, containing no occurrence of the symbols $H_T$, 
has a proof in $HOL^{\pm}$, then it has a proof in $HOL$.
\end{proposition}

\proof{Using Proposition \ref{aaa}, all we need to prove is that the 
theory ${\cal U}_{HOL^{\pm}}$ is a conservative extension of 
${\cal U}_{HOL}$. 

The theories ${\cal U}_{HOL^{\pm}}$ and ${\cal U}_{HOL}$ differ on
three points. First, the theory 
${\cal U}_{HOL^{\pm}}$ contains the axioms 
$\fa x \fa y~(\neg \neg \varepsilon(x) \Rightarrow \varepsilon(x~\dotvee~y))$
and 
$\fa x \fa y~(\neg \neg \varepsilon(y) \Rightarrow \varepsilon(x~\dotvee~y))$
while the theory ${\cal U}_{HOL}$ contains the axiom 
$\fa x \fa y~((\varepsilon(x) \vee \varepsilon(y)) \Rightarrow
\varepsilon(x~\dotvee~y))$. But the conjunction of the two axioms of 
${\cal U}_{HOL^{\pm}}$ is equivalent to that of ${\cal U}_{HOL}$.

Second, the theory ${\cal U}_{HOL}$ contains two axioms 
$\varepsilon(\nulll~0) \Rightarrow \top$ and 
$\fa x~(\bot \Rightarrow \varepsilon(\nulll~(S~x)))$. But these axioms
are trivially provable in predicate logic and they can be eliminated.

Third, the theory ${\cal U}_{HOL^{\pm}}$ contains the axiom 
$\fa x~(\neg \neg \varepsilon(x~H_T(x)) \Rightarrow
\varepsilon(\dotfa_T~x))$
and the axioms of equality for the symbols $H_T$ 
and the theory ${\cal U}_{HOL}$ the axiom 
$\fa x~((\fa y~\varepsilon(x~y)) \Rightarrow \varepsilon(\dotfa_T~x))$. 
But the axiom of ${\cal U}_{HOL^{\pm}}$ is equivalent to the
Skolemization of that of ${\cal U}_{HOL}$. 

Thus, using Skolem theorem for classical logic with equality, we get
that ${\cal U}_{HOL^{\pm}}$ is a conservative extension of 
${\cal U}_{HOL}$.}

\medskip

Trivially, 
if a sequent $\Gamma \vdash \Delta$, containing no occurrence of the symbols 
$H_T$, 
has a cut free proof in $HOL^{\pm}$, it has a proof in $HOL^{\pm}$ and
thus it has a proof in $HOL$. Using the cut elimination theorem for
$HOL$, we get that it has a cut free proof in $HOL$. 
We now want to prove the converse, i.e. that if a sequent,
containing no occurrence of the symbols $H_T$, has a 
cut free proof in $HOL$, it has a cut free proof in $HOL^{\pm}$.

\begin{proposition}
\label{retard}
If $(\Gamma \vdash \Delta) \lra^* (\Gamma' \vdash \Delta')$ and 
$\Gamma' \vdash \Delta'$ has a cut free proof modulo ${\cal R}$ then 
$\Gamma \vdash \Delta$ has a cut free proof modulo ${\cal R}$ of the same size.
\end{proposition}

\proof{By induction over proof structure.}

\begin{proposition}\label{reciproque}
If a sequent containing no occurrence of the symbols $H_T$,
has a cut free proof in $HOL$, it has a cut free proof in $HOL^{\pm}$.
\end{proposition}

\proof{Let $\Gamma \vdash \Delta$ be a sequent that has a cut free
proof in $HOL$. By induction on the size of this proof, 
we build a cut free proof of this sequent in $HOL^{\pm}$.
We give only two cases.
\begin{itemize}
\item If the proof has the form 
$$\irule{\irule{\pi}{\Gamma \vdash B, C, \Delta}{}}
        {\Gamma \vdash A, \Delta}
        {\mbox{$\vee$-right}}$$
with $A \lra^{HOL}_+ (B \vee C)$, then either $A = (B' \vee C')$ or $A$ is 
atomic. 

In the first case we have $B' \lra^{HOL}_+ B$, $C' \lra^{HOL}_+ C$. By
Proposition \ref{retard}, the sequent $\Gamma \vdash B', C', \Delta$ 
has a cut free proof of the same size, by induction hypothesis it has a cut free
proof in 
$HOL^{\pm}$ and we conclude with the $\vee$-right rule. 

In the second, consider a reduction sequence from $A$ to $B \vee C$
and in this reduction sequence, the last atomic proposition $A'$ and
its successor $B' \vee C'$. We have $A \lra_+^{HOL*} A' \lra_+^{HOL}
(B' \vee C') \lra_+^{HOL*} (B \vee C)$. As $A'$ is atomic and $A'
\lra_+^{HOL} (B \vee C)$, we have $A' = \varepsilon(t~\dotvee~u)$, $B'
= \varepsilon(t)$, and $C' = \varepsilon(u)$. As $(\varepsilon(t) \vee
\varepsilon(u)) \lra_+^{HOL*} (B \vee C)$, we have 
$\varepsilon(t) \lra_+^{HOL*} B$ and $\varepsilon(u) \lra_+^{HOL*} C$. 
By Proposition \ref{retard}, the sequent $\Gamma \vdash \varepsilon(t), 
\varepsilon(u), \Delta$ has a cut free proof of the same size in $HOL$ and by 
induction hypothesis, it has a cut free proof in $HOL^{\pm}$. As $A$ and 
$\varepsilon(t~\dotvee~u)$ are atomic and $A \lra_+^{HOL*} 
\varepsilon(t~\dotvee~u)$, we have $A \lra_+^{HOL^{\pm}*} 
\varepsilon(t~\dotvee~u)$. Then, $\varepsilon(t~\dotvee~u) 
\lra_+^{HOL^{\pm}} \neg \neg \varepsilon(t)$ and $\varepsilon(t~\dotvee~u) 
\lra_+^{HOL^{\pm}} \neg \neg \varepsilon(u)$.  Thus,
$A \lra_+^{HOL^{\pm}*} \neg \neg \varepsilon(t)$ and $A \lra_+^{HOL^{\pm}*} 
\neg \neg \varepsilon(u)$. We build a cut free proof of $\Gamma \vdash A, 
\Delta$ in $HOL^{\pm}$ with the rules contraction-right, 
$\neg$-right, and $\neg$-left and the proof of $\Gamma \vdash 
\varepsilon(t), \varepsilon(u), \Delta$.

\item If the proof has the form 
$$\irule{\irule{\pi}{\Gamma \vdash B, \Delta}{}}
        {\Gamma \vdash A, \Delta}
        {\mbox{$\fa$-right}}$$
with $A \lra^{HOL}_+ \fa x~B$, then either $A = \fa x~B'$ or $A$ is 
atomic. 

In the first case we have $B' \lra^{HOL}_+ B$. By
Proposition \ref{retard}, the sequent $\Gamma \vdash B', \Delta$ 
has a cut free proof of the same size, by induction hypothesis it has 
a cut free proof in 
$HOL^{\pm}$ and we conclude with the $\fa$-right rule. 

In the second, consider a reduction sequence from $A$ to $\fa x~B$
and in this reduction sequence, the last atomic proposition $A'$ and
its successor $\fa x~B'$. We have $A \lra_+^{HOL*} A' \lra_+^{HOL}
\fa x~B' \lra_+^{HOL*} \fa x~B$. As $A'$ is atomic and $A'
\lra_+^{HOL} \fa x~B$, we have $A' = \varepsilon(\dotfa_T~t)$ and $B'
= \varepsilon(t~x)$. As $\fa x~\varepsilon(t~x) 
\lra_+^{HOL*} \fa x~B$, we have 
$\varepsilon(t~x) \lra_+^{HOL*} B$. 
By Proposition \ref{retard}, the sequent $\Gamma \vdash \varepsilon(t~x), 
\Delta$ has a cut free proof of the same size in $HOL$ and by 
induction hypothesis, it has a cut free proof in $HOL^{\pm}$. 
By substituting the term $H_T(t)$ for the variable $x$ in this 
proof, we get a proof of the sequent $\Gamma \vdash \varepsilon(t~H_T(t))$ 
in $HOL^{\pm}$. 
As $A$ and $\varepsilon(\dotfa_T~t)$ are atomic and $A \lra_+^{HOL*} 
\varepsilon(\dotfa_T~t)$, we have $A \lra_+^{HOL^{\pm}*} 
\varepsilon(\dotfa_T~t)$. Then, $\varepsilon(\dotfa_T~t) 
\lra_+^{HOL^{\pm}} \neg \neg \varepsilon(t~H_T(t))$.  
Thus, $A \lra_+^{HOL^{\pm}*} \neg \neg \varepsilon(t~H_T(t))$. 
We build a cut free proof of $\Gamma \vdash A, \Delta$ in $HOL^{\pm}$ with the 
rules $\neg$-right and $\neg$-left and the proof of $\Gamma \vdash 
\varepsilon(t~H_T(t)), \Delta$.
\end{itemize}}

\begin{proposition}
For a sequent containing no occurrence of the symbols $H_T$ the
following conditions are equivalent 
\begin{enumerate}
\item the sequent has a proof in $HOL^{\pm}$, 
\item it has a proof in $HOL$,
\item it has a cut free proof in $HOL$,
\item it has a cut free proof in $HOL^{\pm}$.
\end{enumerate}
\end{proposition}

\proof{{\em 1.} $\Rightarrow$ {\em 2.} is Proposition \ref{sensdirect}, 
{\em 2.} $\Rightarrow$ {\em 3.} is the cut elimination for $HOL$ 
\cite{Prawitz,Takahashi,Girard} (see also \cite{DowekWerner}),
{\em 3.} $\Rightarrow$ {\em 4.} is Proposition \ref{reciproque}, 
{\em 4.} $\Rightarrow$ {\em 1.} is trivial.}

\medskip

Notice that $HOL^{\pm}$ does not have the cut elimination property in 
general. For instance, the sequent $\varepsilon(x~H_T(x)) \vdash 
\fa y~\varepsilon(x~y)$ has a proof with a cut 
(on $\varepsilon(\dotfa_T~x)$) but no cut free proof. 
Yet, for sequents well-formed in the language of $HOL$ (i.e. containing 
no symbols $H_T$), the cut elimination property holds and provability 
is equivalent to provability in $HOL$.


\begin{thebibliography}{99.}
\bibitem{stacs} 
G. Dowek, What is a theory?, H. Alt, A. Ferreira (Eds.), {\em
Symposium on Theoretical Aspects of Computer 
Science}, Lecture Notes in Computer Science 2285, Springer-Verlag,
2002, pp. 50-64. 

\bibitem{polar} 
G. Dowek, Polarized Resolution Modulo, manuscript, 2009.

\bibitem{DHKHOL} 
G. Dowek, Th. Hardin, and C. Kirchner, 
HOL-lambda-sigma: an intentional first-order expression of
higher-order logic, {\em Mathematical Structures in Computer Science}, 11,
2001, pp. 1-25.  

\bibitem{DowekWerner} 
G. Dowek and B. Werner, Proof normalization modulo, {\em The Journal of
Symbolic Logic}, 68, 4, 2003, pp. 1289-1316.  

\bibitem{Girard} 
J.-Y. Girard, Une extension de l'interpr\'etation de G\"odel \`a
l'analyse et son application \`a l'\'elimination des coupures dans
l'analyse et la th\'eorie des types, J.E.~Fenstad (Ed.)  {\em Second
Scandinavian Logic Symposium}, North-Holland, 1970.

\bibitem{Prawitz} 
D.~Prawitz. Hauptsatz for higher order logic. {\em The Journal of
Symbolic Logic}, 33:452--457, 1968.

\bibitem{Takahashi}
M.~o. Takahashi. A proof of cut-elimination theorem in simple type theory.
{\em Journal of the Mathematical Society of Japan}, 19:399--410, 1967.
\end{thebibliography}
\end{document}